\documentclass[twocolumn,prb]{revtex4}
\usepackage{graphicx}
\usepackage{dcolumn}
\usepackage{bm}
\usepackage{amsmath}

\begin{document}

\preprint{}
\title{A quantitative model for $I_{\rm C}R$ product in $d$-wave Josephson junctions}
\author{T. Yokoyama$^{1,2}$, Y. Sawa$^{1,2}$, Y. Tanaka$^{1,2}$, A. A. Golubov$^3$, A. Maeda$^{2,4}$ and A. Fujimaki$^{2,5}$}
\affiliation{$^1$Department of Applied Physics, Nagoya University, Nagoya, 464-8603, Japan\\%
$^2$ CREST, Japan Science and Technology Corporation (JST) Nagoya, 464-8603,
Japan \\
$^3$ Faculty of Science and Technology, University of Twente, 7500 AE,
Enschede, The Netherlands\\
$^4$ Department of Basic Science, University of Tokyo, 3-8-1, Komaba, Meguro-ku, Tokyo, 153-8902, Japan\\
$^5$Department of Quantum Engineering, Nagoya University, Nagoya, 464-8603, Japan%
}
\date{\today}

\begin{abstract}
We study theoretically the Josephson effect in $d$-wave superconductor / diffusive normal metal /insulator/ diffusive normal metal/ $d$-wave
superconductor (D/DN/I/DN/D) junctions. This model is aimed to describe practical junctions in high-$T_C$ cuprate superconductors, in which the product of the critical Josephson current ($I_C$) and the normal state resistance ($R$) (the so-called $I_{\rm C}R$ product) is very small compared to the prediction of the standard theory.
We show that the $I_{\rm C}R$ product in D/DN/I/DN/D junctions can be much smaller than that in  $d$-wave superconductor / insulator / $d$-wave superconductor junctions and formulate the conditions necessary to achieve large $I_{\rm C}R$ product in D/DN/I/DN/D junctions.
The proposed theory describes the behavior of $I_{\rm C}R$ products quantitatively in high-$T_{\rm C}$ cuprate junctions.
\end{abstract}

\pacs{PACS numbers: 74.20.Rp, 74.50.+r, 74.70.Kn}
\maketitle




%

%




Josephson effect in high $T_{\rm C}$ superconductors has attracted much attention\cite{Tsuei,Harlin} because of its potential applications in future technologies\cite{Hilgenkamp}.
In particular, applications in electronics, such as the single-flux quantum devices are extremely promising, since the operating frequency
is proportional to the product of the critical Josephson current ($I_C$) and the normal state resistance ($R$) (the so-called $I_{\rm C}R$ product) which is approximately proportional to the superconductivity critical temperature\cite{Josephson}.
However, almost all experimental data for high-$T_{\rm C}$ Josephson junctions fabricated so far have shown  $I_{\rm C}R$ values much smaller than those predicted by the standard theory, irrespective of what kind of junctions they are \cite{Delin,Yoshida}.
This strongly suggests that the interfaces in the cuprates are intrinsically pair breaking. Several models were proposed which treat this issue \cite{Deutscher,Gross,Halbritter}. However, the situation was not described so far in a quantitative manner.

\begin{figure}[htb]
\begin{center}
\scalebox{0.4}{
\includegraphics[width=17.0cm,clip]{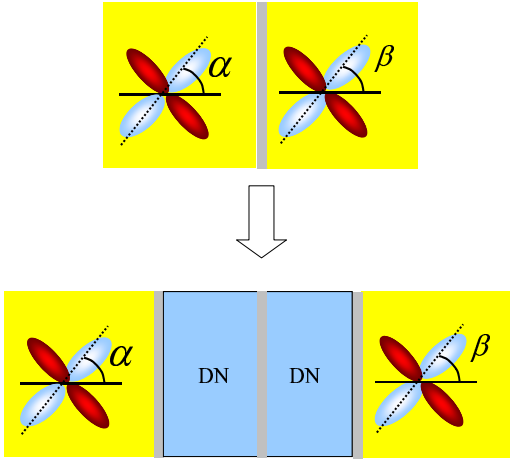}}
\end{center}
\caption{ (color online) Schematic illustration of the model for the D/DN/I/DN/D junction.}
\label{f1}
\end{figure}

One possibility addressed in the present paper and not investigated before is that superconductivity is destroyed near the interface in $d$-wave superconductor / insulator / $d$-wave superconductor (DID) junctions, where diffusive normal metal (DN) regions are induced.
Thus, DID junctions turn into
$d$-wave superconductor / diffusive normal metal /insulator/ diffusive normal metal/ $d$-wave
superconductor (D/DN/I/DN/D) junctions (see Fig. \ref{f1}). In these junctions, $I_C R$ product can be much smaller than that in DID junctions.
In this paper we will explore this possibility and provide a quantitative model which is compared to the experimental data.

The Josephson effect is a phase-sensitive phenomenon and thus depends strongly on a superconducting pairing symmetry \cite{Tsuei,Harlin,SR95}. In DID junctions, nonmonotonic temperature dependence of critical current\cite{Barash,TK96,Golubov2,Ilichev,Testa} occurs due to the formation of midgap
Andreev resonant states (MARS) at the interface\cite{Buch}. The MARS stem from sign change of pair potentials of $d$-wave superconductors \cite%
{Tanaka95}. It was also predicted that MARS strongly enhance the Josephson current at low temperatures\cite{TK96}. On the other hand, in Josephson junctions with DN, the role of the MARS change.

In superconductor / diffusive normal metal / superconductor (S/DN/S) junctions Cooper pairs penetrate into the DN as a result of the proximity effect, providing the Josephson coupling\cite{Likharev,Golubov,Zaikin,Kupriyanov,Zaitsev}. Scattering of electrons by impurities in the DN layer makes superconducting coherence length shorter and thus suppresses the Josephson current. In D/DN/D junctions, the Josephson current is suppressed by the MARS\cite{Asano,Yokoyama,Yokoyama2,Yokoyama3}, in contrast to DID junctions, because MARS compete with proximity effect\cite{Nazarov2003,TNGK}.
Therefore $I_C R$ product in D/DN/I/DN/D junctions can be much smaller than that in DID junctions.

In the present paper, we calculate Josephson current in D/DN/I/DN/D junctions as a model of the actual DID (e.g., grain boundary) junctions.  We show that $I_C R$ product in D/DN/I/DN/D junctions can be much smaller than that in DID junctions, and clarify
the conditions with which the $I_{\rm C}R$ product is most enhanced in D/DN/I/DN/D junctions.
Our theory can explain the above mentioned general trend of the high-$T_C$ Josephson junctions  quantitatively, in contrast to previous theoretical models of high-$T_{\rm C}$ cuprate junctions.
The obtained results may provide useful information for fabrication of high-$T_C$ Josephson junctions.

Let us formulate the model for a D/DN/I/DN/D junction.
We assume that the DN layer has a length $L$ much larger than the mean
free path and is characterized by the resistance $R_{d}$. The DN/D interfaces located at $x=\pm L$ have the resistance 
$R_{b}^{\prime }$, while the DN/I interface at $x=0$ has the
resistance $R_{b}$. We model infinitely narrow insulating barriers  
by the delta function $U(x)=H^{\prime }\delta (x+L)+H\delta
(x)+H^{\prime }\delta
(x-L)$. The resulting transparencies of the interfaces $T_{m}$ and $%
T_{m}^{\prime }$ are given by $T_{m}=4\cos ^{2}\phi /(4\cos ^{2}\phi +Z^{2})$
and $T_{m}^{\prime }=4\cos ^{2}\phi /(4\cos ^{2}\phi +{Z^{\prime }}^{2})$,
where $Z=2H/v_{F}$ and $Z^{\prime }=2H^{\prime }/v_{F}$ are dimensionless
constants and $v_{F}$ is Fermi velocity, where
$\phi $ is the injection angle measured from the interface normal. In the following we assume $Z \gg 1$.
The schematic illustration of the model is shown in Fig. \ref{f1}. The pair potential along the quasiparticle trajectory
with the injection angle $\phi $ is given by $\Delta _{L}=\Delta \cos
[2(\phi -\alpha )]\exp (-i\Psi )$ and $\Delta _{R}=\Delta \cos [2(\phi
-\beta )]$ for the left and the right superconductors, respectively. Here
$\Psi$ is the phase difference across the junction, $\alpha $ and $\beta $ denote the
angles between the normal to the interface and the crystal axes of the left
and right $d$-wave superconductors, respectively. The lobe
direction of the pair potential and the direction of the crystal axis are
chosen to be the same.

We parameterize the quasiclassical Green's functions $G$ and $F$
with a function $\Phi _{\omega }$ \cite{Likharev,Golubov}:
\begin{equation}
G_{\omega }=\frac{\omega }{\sqrt{\omega ^{2}+\Phi _{\omega }\Phi _{-\omega
}^{\ast }}},F_{\omega }=\frac{{\Phi _{\omega }}}{\sqrt{\omega ^{2}+\Phi
_{\omega }\Phi _{-\omega }^{\ast }}}
\end{equation}%
where $\omega $ is the Matsubara frequency. In the DN layers the Green's functions 
satisfy the Usadel equation
\cite{Usadel}
\begin{equation}
\xi ^{2}\frac{{\pi T_{C}}}{{\omega G_{\omega }}}\frac{\partial }{{\partial x}
}\left( {G_{\omega }^{2}\frac{\partial }{{\partial x}}\Phi _{\omega }}
\right) -\Phi _{\omega }=0
\end{equation}%
where $\xi =\sqrt{D/2\pi T_{C}}$ is the coherence length, $D$ is the diffusion constant
and $T_{C}$ is the transition temperature of superconducting electrodes. To solve the Usadel equation,
we apply the generalized boundary conditions derived in Ref.\cite{Yokoyama,Yokoyama2} at $x= \pm L$ and 
the boundary conditions in Ref.\cite{Kupriyanov} at $x=0$.

The Josephson current is given by
\begin{equation}
\frac{{eIR}}{{\pi T_C }} = i\frac{{RTL}}{{2R_d T_C }}\sum\limits_\omega {%
\frac{{G_\omega ^2 }}{{\omega ^2 }}} \left( {\Phi _\omega \frac{\partial }{{%
\partial x}}\Phi _{ - \omega }^ * - \Phi _{ - \omega }^ * \frac{\partial }{{%
\partial x}}\Phi _\omega } \right)
\end{equation}
where $T$ is temperature and $R \equiv 2R_{d}+R_{b}+2R_{b}^{\prime}$ is the
normal state resistance of the junction. In the following we focus on the $I_C R$ value as a function of temperature 
and clarify the cases when $I_C R$ is enhanced. Below $\Delta(0)$ denotes the value of $\Delta$ at zero temperature. Note that it is realistic to choose small magnitude of $Z^{\prime}$ and $ R_{b}^{\prime}$,  and large Thouless energy because thin DN regions could be naturally formed due to the
degradation of superconductivity near the interface.


\begin{figure}[tbh]
\begin{center}
\scalebox{0.4}{
\includegraphics[width=18.0cm,clip]{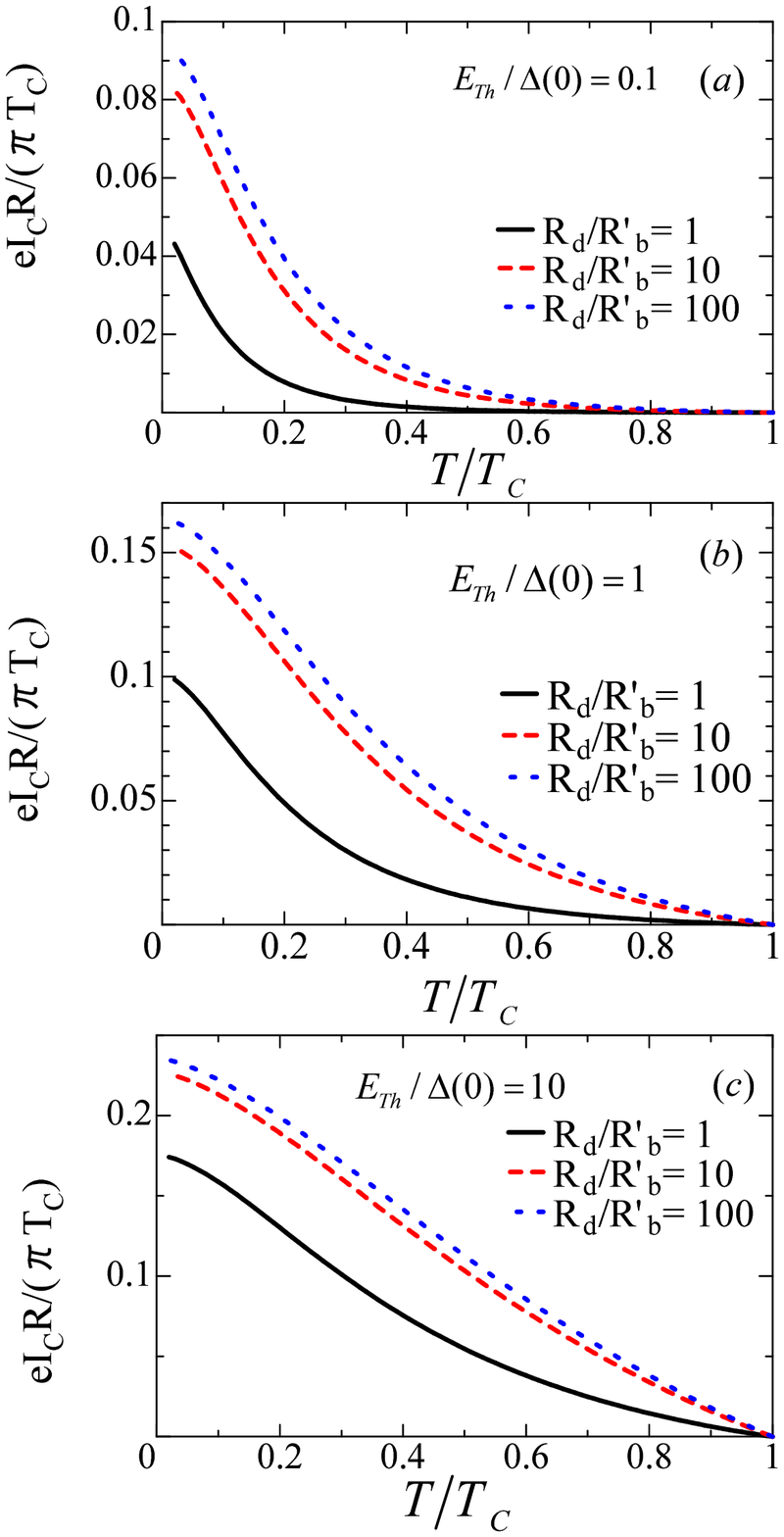}}
\end{center}
\caption{(color online) $I_{C}R$ value  for $Z^{\prime }=1$, $R_{d}/R_b=0.1$  and $\left( {%
\alpha ,\beta }\right) =\left( {0,0}\right) $.}
\label{f2}
\end{figure}

\begin{figure}[tbh]
\begin{center}
\scalebox{0.4}{
\includegraphics[width=16.0cm,clip]{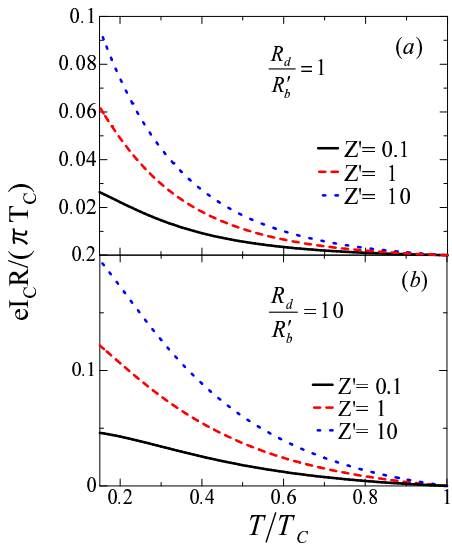}}
\end{center}
\caption{(color online) $I_{C}R$ value for $E_{Th}/\Delta (0)=1$, $R_{d}/R_b=0.1$  and $\left( {%
\alpha ,\beta }\right) =\left( {0,0}\right) $.}
\label{f3}
\end{figure}

In Fig.\ref{f2} we show $I_{C}R$ value for $Z^{\prime }=1$, $R_{d}/R_b=0.1$  and $\left( {%
\alpha ,\beta }\right) =\left( {0,0}\right) $ with various $E_{Th}/\Delta (0)$ and $R_{d}/R_{b}^{\prime}$. $I_{C}R$ increases with $E_{Th}/\Delta (0)$ and $R_{d}/R_{b}^{\prime}$ because proximity effect is enhanced. As $E_{Th}$ increases, the magnitude of the gradient becomes small.

Figure \ref{f3} shows $I_{C}R$ value for $E_{Th}/\Delta (0)=1$, $R_{d}/R_b=0.1$  and $\left( {%
\alpha ,\beta }\right) =\left( {0,0}\right) $ with various $Z^{\prime }$ and $R_{d}/R_{b}^{\prime}$.  As $Z^{\prime }$ increases, the magnitude of the gradient becomes large. The peculiar effect is that $I_{C}R$ increases with $Z^{\prime }$, indicating that proximity effect is enhanced by the increase of $Z^{\prime }$. This stems from the sign change of the pair
potential\cite{Yokoyama,Yokoyama2}.  For the case of $d$-wave symmetry with $\alpha=\beta=0$, injection angles of a quasiparticle
can be separated into two regions: $\phi_+= \{ \phi | 0 \leq |\phi|< \pi/4 \}$
and $\phi_- =\{ \phi | \pi/4 \leq |\phi| \leq \pi/2 \} $. The signs of pair potential for $\phi_+$ and that for $\phi_-$ are opposite.
As a result, the sign change of pair potentials
suppresses the proximity effect in the DN and hence Josephson currents. As $Z^{\prime }$ increases, the contribution from $\phi_+$ dominates 
over that from $\phi_-$. Therefore $I_{C}R$ increases with $Z^{\prime }$.

\begin{figure}[tbh]
\begin{center}
\scalebox{0.4}{
\includegraphics[width=17.0cm,clip]{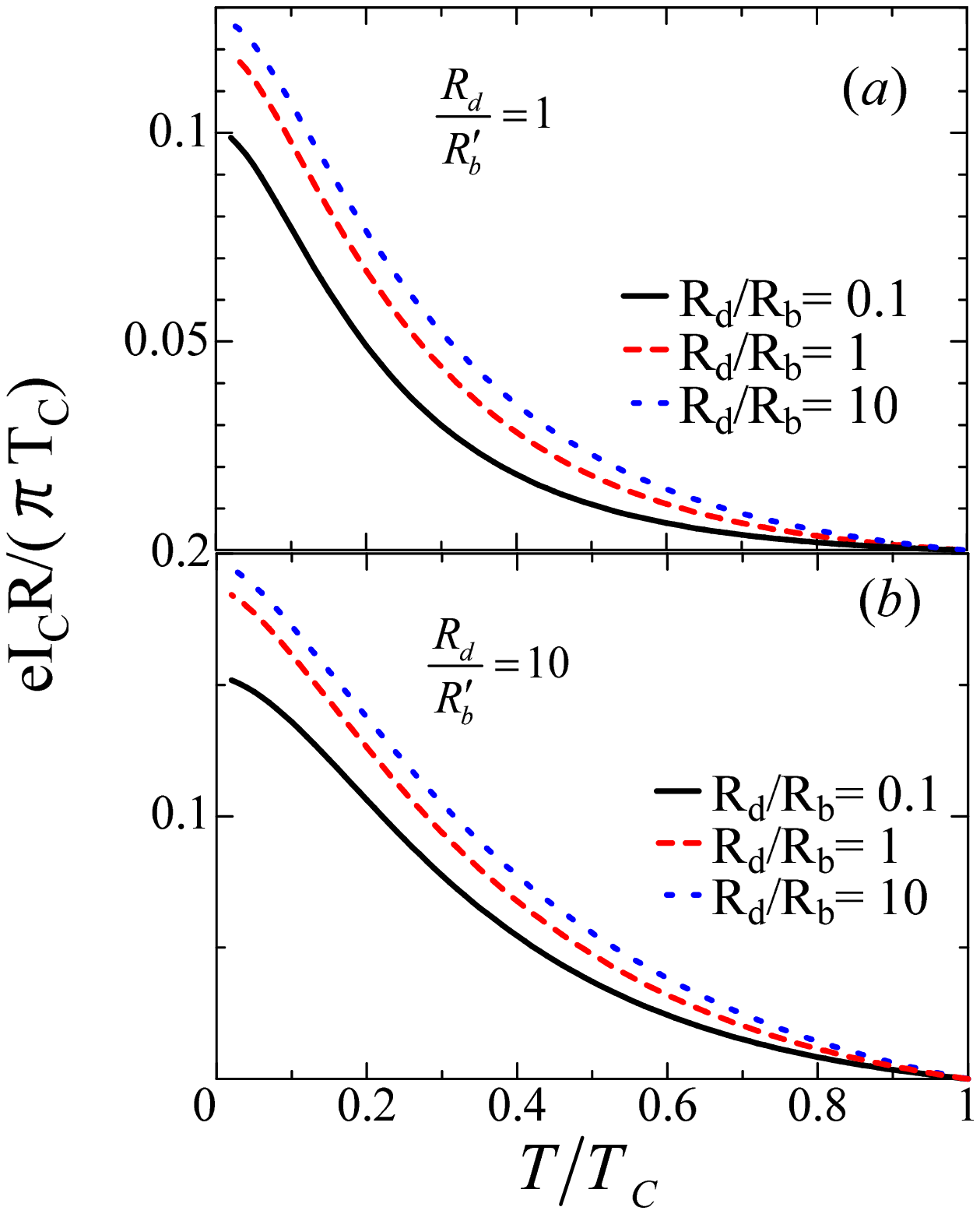}}
\end{center}
\caption{(color online) $I_{C}R$ value for $E_{Th}/\Delta (0)=1$, $Z^{\prime }=1$  and $\left( {%
\alpha ,\beta }\right) =\left( {0,0}\right) $.}
\label{f4}
\end{figure}

\begin{figure}[tbh]
\begin{center}
\scalebox{0.4}{
\includegraphics[width=16.0cm,clip]{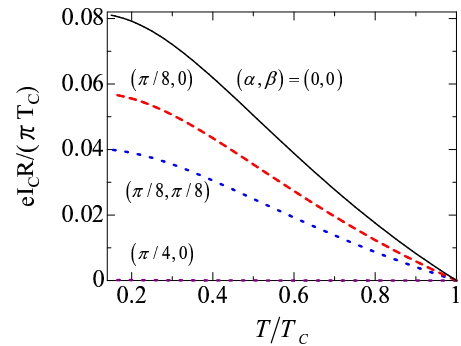}}
\end{center}
\caption{(color online) $I_{C}R$ value for $E_{Th}/\Delta (0)=0.1$, $Z^{\prime }=0.1$, $R_{d}/R_b=0.1$ and $R_{d}/R_{b}^{\prime}=10$.}
\label{f5}
\end{figure}

\begin{figure}[tbh]
\begin{center}
\scalebox{0.4}{
\includegraphics[width=16.0cm,clip]{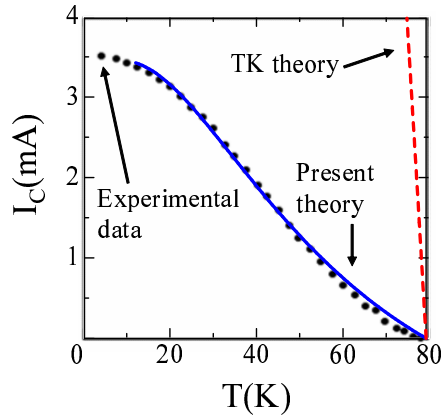}}
\end{center}
\caption{(color online) Comparison between the present theory (solid line), experimental data\cite{Yoshida}  (dotted line) and TK theory\cite{TK96} (broken line). $I_{C}$  is plotted as a function of temperature, taking $R=0.375\Omega$ and $\alpha=\beta=0$ for theoretical plots.
We choose $E_{Th}/\Delta (0)=3$, $Z^{\prime}=0.1$, $R_{d}/R_b=0.01$, and  $R_{d}/R_{b}^{\prime}=100$ in the present theory, and $Z=10$ in TK theory.}
\label{f6}
\end{figure}

In Fig. \ref{f4} we plot $I_{C}R$ value  for $E_{Th}/\Delta (0)=1$, $Z^{\prime }=1$  and $\left( {%
\alpha ,\beta }\right) =\left( {0,0}\right) $ with various $R_{d}/R_b$ and $R_{d}/R_{b}^{\prime}$. $I_{C}R$ increases with $R_{d}/R_b$  
due to the enhancement of the proximity effect.

Figure \ref{f5} displays $I_{C}R$ value for $E_{Th}/\Delta (0)=0.1$, $Z^{\prime }=0.1$, $R_{d}/R_b=0.1$ and $R_{d}/R_{b}^{\prime}=10$  with various $\alpha$ and $\beta$. The formation of MARS suppresses the proximity effect. Therefore $I_{C}R$ decreases with the increase of $\alpha$ and $\beta$\cite{Asano,Yokoyama,Yokoyama2,Yokoyama3}.
In the actual junctions, there is inevitable roughness at the interface and hence the effective values of $\alpha$ and $\beta$ at the interface become random even if junctions with $\alpha=\beta=0$ are fabricated. This provides the mechanism of suppression of the $I_{C}R$ product.

Finally we compare the present theory with the experimental data from Ref.\cite{Yoshida} and with the theory for DID junctions by Tanaka and 
Kashiwaya (TK)\cite{TK96}. The temperature dependencies of $I_{C}$ are plotted in Fig. \ref{f6} taking $\alpha=\beta=0$ and $R=0.375\Omega$
for theoretical plots.
We choose $E_{Th}/\Delta (0)=3$, $Z^{\prime}=0.1$, $R_{d}/R_b=0.01$, and  $R_{d}/R_{b}^{\prime}=100$ in the present theory, and the barrier parameter $Z=10$ in the TK theory. As shown in this figure, the present theory can explain the experimental results quantitatively, while the discrepancy between the TK theory and the data is rather strong, about an order of magnitude. Note that in the TK theory the $I_{C}$ is not sensitive to the choice of $Z$ parameter. To estimate the realistic size of the DN region, we can take $\Delta (0)=10$meV and $D=10^{-3}m^{2}/s$, and then obtain the length of the DN region $L=4.7$nm.

In summary, we have studied the Josephson current in D/DN/I/DN/D junctions as a model of high $T_C$ superconductor junctions. We have shown that the $I_C R$ product in D/DN/I/DN/D junctions can be much smaller than that in DID junctions and have found the conditions when the $I_C R$ in D/DN/I/DN/D junctions is largest. The requirements for the large magnitude of $I_C R$ product are: no roughness at the interfaces, large magnitudes of  $Z^{\prime }$, $R_{d}/R_{b}$, $R_{d}/R_{b}^{\prime }$ and $E_{Th}$, and $\left( {\alpha ,\beta }\right) =\left( {0,0}\right) $. Note that small magnitude of $Z^{\prime}$ and $ R_{b}^{\prime}$, and large $E_{Th}$ are realistic for naturally formed DN layers, hence the only tunable parameter is $R_{b}$. Our theory can explain the experimental results on the quantitative level, in contrast to the previous idealized treatment of DID junctions.

T. Y. acknowledges support by the JSPS. This work is supported by Grant-in-Aid for Scientific Research
on Priority Area "Novel Quantum Phenomena Specific to Anisotropic
Superconductivity" (Grant No. 17071007) from the Ministry of Education,
Culture, Sports, Science and Technology of Japan, and also supported by
NAREGI Nanoscience Project, the Ministry of Education, Culture,
Sports, Science and Technology, Japan, the Core Research for Evolutional
Science and Technology (CREST) of the Japan Science and Technology
Corporation (JST), a Grant-in-Aid for the 21st Century COE "Frontiers of
Computational Science" and NanoNed Project TCS7029. The computational aspect of this work has been
performed at the Research Center for Computational Science, Okazaki National
Research Institutes and the facilities of the Supercomputer Center,
Institute for Solid State Physics, University of Tokyo and the Computer Center.
%


\end{document}